\documentstyle[epsf,12pt]{article}
\newcommand{\gsim}{\mbox{\raisebox{-.3em}{$\stackrel{>}{\sim}$}}}
\newcommand{\lsim}{\mbox{\raisebox{-.3em}{$\stackrel{<}{\sim}$}}}

\renewcommand{\cite}[1]{\ref{#1}}
\newcommand{\BBox}{\mbox{\raisebox{-0.2em}{\large$\Box$}}}
\newcommand{\half}{\frac{1}{2}}

\newcommand{\psibar}{\overline{\psi}}
\newcommand{\Dslash}{D\hspace{-.65em}/}

\newcommand{\psls}{p\hspace{-.4em}/}
\newcommand{\beq}{\begin{equation}}
\newcommand{\eeq}{\end{equation}}
\newcommand{\beqa}{\begin{eqnarray}}
\newcommand{\eeqa}{\end{eqnarray}}

\newcommand{\reflef}{(\ref}

\begin{document}
%\begin{flushright}
%bd10bt1.tex\\
%\today
%\end{flushright}
%\baselineskip=0.7cm
\begin{center}
{\large\bf Brans-Dicke Cosmology Corrected for a Quantum \vspace{.2em}Effect due to the Scalar-Matter Coupling}\vspace{.3cm}\\
Yasunori Fujii\footnote{Electronic address: fujii@handy.n-fukushi.ac.jp}\\
Nihon Fukushi University, Handa, 475\ Japan\\
and\\
ICRR, University of Tokyo, Tanashi, Tokyo, 188 Japan \vspace{.5cm}\\
{\bf abstract}
\end{center}
Cosmological solutions of the Brans-Dicke theory are investigated by
including a quantum effect coming from 1-loop correction of matter
fields that couple to the scalar field.  As the most serious result we 
face a cosmological ``constant'' in the original conformal frame which
is shown to be ``physical'' after a careful analysis of the time
variable employed in any of the conventional approaches.  We find an
``attractor'' solution featuring no expansion during the
radiation-dominated eras.  To evade this unacceptable consequence, we
suggest to modify one of the fundamental premises of the model,
rendering the scalar field almost ``invisible.''\\

\section{Introduction}
Among many versions of the scalar-tensor theory of gravitation, the
prototype Brans-Dicke (BD) model [\cite{bd}] is unique for the
following four assumptions on the scalar field: (i) a nonminimal coupling of the simplest form; (ii) masslessness; (iii) no self-interaction; (iv) no direct matter coupling.

Although the model may not be fully realistic, it still seems to deserve further scrutiny as a testing ground of many aspects of wider class of the scalar-tensor theories.  The model has been studied,
 however, mainly as a classical theory.  We attempt to take  
quantum effects due to the matter coupling into account.  It has
been argued that composition-independence that entails from the
assumption (iv) above would be violated as a quantum correction [\cite{yf1}].  We
 came to realize, however, that the suspected contribution is canceled by other 
terms arising from regularization;\footnote{More details on this analysis will be reported in another publication.} WEP is in fact a well-protected
and robust property of the
model beyond the classical level.

We discuss in this note another quantum effect which, as it turns out, has serious
cosmological consequences, beyond the extent of remedy expected by adjusting
the fundamental parameter of the model.  For a possible way out we suggest to modify the assumption
(iv), making the scalar field almost ``invisible," thus reconciling with the absence of experimental evidences, still playing a cosmological role.  We also discuss how the scalar field acquires a nonzero
self-mass due to the matter coupling, even having started with 
the {\em classical} 
assumption (ii), but likely in an entirely insignificant manner in practice.

We start with the basic Lagrangian
\beq
{\cal L}=\sqrt{-g}\left( \half F(\phi) R 
-\epsilon\half g^{\mu\nu}\partial_{\mu}\phi \partial_{\nu}\phi 
+L_{\rm m}\right),
\label{bd5_51}
\eeq
with
\beq
F(\phi)=\xi\phi^2.
\label{bd5_52}
\eeq
We use the unit system of $8\pi G=1.$\footnote{The units of
length, time and energy are $8.07\times 10^{-33}$cm, $2.71\times
10^{-43}$ sec, and $2.43\times 10^{18}$ GeV, respectively.  Notice
also that the present age of the Universe is $\sim 10^{60}$.}  The
scalar field $\phi$ and the constant $\xi$ are related to the original notation $\varphi$ and $\omega$, respectively, by 
\beq
\varphi =\half\xi\phi^2, \quad{\rm and }\quad \omega =\frac{1}{4\xi}.
\label{bd5_53}
\eeq
We also allow $\epsilon =\pm 1$, a minimum extension of the original model to avoid an immediate failure.  As we see shortly, $\epsilon =-1$ does
not necessarily imply a ghost in the final result.

As the matter Lagrangian we choose, according to the assumption (iv),
\beq
L_{\rm m}= -\psibar \left( \Dslash +m_{0} \right)\psi,
\label{bd55_4}
\eeq
where $\psi$ stands for a simplified representative of the (spinor) matter
field.

In spite of (iv) $\phi$ couples in effect to the matter field {\em in the field equation.}  This is inconvenient, however, when we try to apply the conventional technique of quantum field theory to the $\phi\,$-matter coupling.  For this reason we apply a conformal transformation such that the nonminimal coupling is eliminated [\cite{dicke}]:
\beq
g_{\mu\nu}=\Omega^{-2}g_{*\mu\nu},\quad{\rm with}\quad \Omega =\left(\xi\phi^2\right)^{1/2}.
\label{bd55_5}
\eeq
Notice that we should have 
\beq
\xi >0, 
\label{bd55_5a}
\eeq
in order to ensure that the sign of the line element remains unchanged.

We in fact find that (\ref{bd5_51}) is now put into the form
\beq
{\cal L}=\sqrt{-g_{*}}\left( \half R_{*} -\half g^{\mu\nu}_{*}\partial_{\mu}\sigma\partial_{\nu}\sigma
   +L_{*m}  \right),
\label{bd3-16}
\eeq
expressed in terms of the new metric $g_{*\mu\nu}$ and the field $\psi_{*}=\Omega^{-3/2}\psi$.  Also the {\em canonical} scalar field is now $\sigma$ as defined by
\beq
\phi(\sigma)=\phi_{0}\;e^{\beta\sigma},
\label{bd5-2}
\eeq
where
\beq
\beta =\frac{1}{\sqrt{6+\epsilon\xi^{-1}}}
=\sqrt{\frac{4\pi G}{3+2\epsilon\omega}}.
\label{bd5-2a}
\eeq
We emphasize that $\sigma$ is not a ghost if 
\beq
\beta^2 >0,
\label{bd5-2b}
\eeq
even if $\epsilon =-1$ [\cite{fn}]; ``mixing" between the scalar field and (spinless part of) the metric field provides sufficient amount of positive contribution to overcome the negative kinetic part.  The condition (\ref{bd5-2b}) is
equivalent to
\beq
\epsilon\xi^{-1} > -6,\quad {\rm or}\quad \epsilon\omega >-\frac{3}{2}.
\label{bd5-2c}
\eeq

We now have a direct $\sigma$-matter coupling expressed in terms of
the interaction Hamiltonian to which usual perturbation method is
readily applied.  We point out, however, that EEP, hence WEP as well, remains intact because the deviation from geodesic arising from the transformation (\ref{bd55_5}) is given entirely in terms of $\sigma$ which is independent of any specific properties of individual particles; the fact that any motion is independent of the mass, for example, verified in one conformal frame obviously survives conformal transformations.

We call the conformal frames before and after the conformal
transformation J frame and E frame, respectively.\footnote{The name J
frame is used, following Cho's suggestion [\cite{cho}], after
 P. Jordan [\cite{jd}] who was the first to discuss the nonminimal coupling.  On the other hand, E frame is a reminder that this is a frame in which the standard theory of Einstein is formulated.}

\section{Quantum effect}
In E frame, in which we hereafter suppress the symbol $*$ for simplicity, we consider one-loop diagrams as shown in Fig. 1, due to the interaction
\beq
H'_{1}=m_{0}\xi^{-1/2}\phi^{-1}(\sigma)\psibar\psi,
\label{bd5-1}
\eeq
coming originally from the mass term in J frame.

As will be shown, $\sigma$ may keep moving with time, and so does the mass.  But $\sigma$ moves so slowly compared with any of the microscopic time-scales, that the mass $m(t)$ {\em at each epoch} will be defined by 
\beq
m(t)=m_{0}\xi^{-1/2}\phi^{-1}(\sigma(t)).
\label{bd5-3}
\eeq

Now consider the 1\hspace{.2em}-$\phi^{-1}$ diagram (a).  Its contribution is given by the potential of $\sigma$;
\beq
V_{1}(\sigma)=im(\sigma)\int d^4 p \,{\rm Tr}\!
	\left(\frac{1}{m(\sigma)+i\psls}\right).
\label{bd5-4}
\eeq
Our consideration will be restricted to the Universe which is
sufficiently late to justify to ignore the effect of temperature and
spacetime curvature.\footnote{The temperature will be lower than $\sim 
$TeV, for example, if $t\gsim 10^{-13}$sec, much earlier than the epoch of nucleosynthesis.  Spacetime curvature will be important only for $t\lsim 10^{-28}$sec, corresponding to the temperature of $10^{11}$GeV.}

The integral in \reflef{bd5-4}) is quadratically divergent, but is expected to vanish if there is supersymmetry because the fermionic contribution given by (\ref{bd5-4}) is canceled by the same contribution from the bosonic partner.  The cancellation would not be complete, however, if supersymmetry is broken at the mass scale
\beq
M_{\rm ssb}=rm,
\label{bd5-5}
\eeq
where the ratio $r$ of $M_{\rm ssb}$ to $m$, a representative of
ordinary particles taken roughly of the order of GeV, would be $\sim
10^3$--$10^4$, which we naturally choose to be a true constant.

 The result may be given by
\beq
V_{1}(\sigma)=C_{1}r^2 m^4 
	=V_{1}(0)e^{-4\beta\sigma},
\label{bd5-6}
\eeq
where
\beq
V_{1}(0)=C_{1}r^2\left( m_{0}\xi^{-1/2}\phi_{0}^{-1} \right)^4,
\label{bd5-6a}
\eeq
with $C_{1}$ most likely of the order 1.  Its sign, however, may not be known precisely because it depends on the details of supersymmetry breaking.  If $C_{1}>0$, (\ref{bd5-6}) gives a positive exponential potential that would drive $\sigma$ toward infinity.  We assume this to be the case.

\section{Cosmology}
We now consider the cosmological equations with a classical
potential $V=V_{1}$ which is the only potential \vspace{.3em}due to the assumption (iii):
\beqa
&&3H^2 =\half \dot{\sigma}^2+V +\rho, \label{bd5-7}\vspace{.5em}\\
&&\ddot{\sigma}+3H\dot{\sigma}+V'(\sigma)=0,\vspace{.3em}\label{bd5-8}\\
&&\dot{\rho}+4H\rho =0,\label{bd5-9}
\eeqa
where $\rho$ is the matter density which we assume, for the moment, to
be relativistic.  We also ignored possible terms representing the
coupling between $\sigma$ and $\rho$.  This would be justified for our
purposes as long as we consider late epochs during which the coupling
is sufficiently weak [\cite{fn}].

With $V=V_{1}$, a set of analytic solutions of (\ref{bd5-7})--(\ref{bd5-9}) are obtained:
\beqa
a(t)&=& t^{1/2},\label{bd5-10}\\
\sigma(t)&=&\frac{1}{2\beta}\ln t,\label{bd5-11}\\
\rho(t)&=& \frac{3}{4}\left( 1- \frac{1}{4}\beta^{-2} \right)t^{-2},
\label{bd5-12}
\eeqa
with
\beq
V_{1}(0)=\frac{1}{16\beta^2}.
\label{bd5-12a}
\eeq

Notice that the condition $\rho >0$ is met if
\beq
\beta^{-2}<4,\quad {\rm or}\quad \epsilon\xi^{-1}<-2,\label{bd5-13}
\eeq
which, combined with (\ref{bd55_5a}), is satisfied only if $\epsilon =-1$.  This is the reason why we decided to allow an ``apparent" ghost in J frame.  Then (\ref{bd5-13}) translates into 
\beq
\xi^{-1}>2,\quad{\rm or}\quad \omega > \half,
\label{bd5-13a}
\eeq
a much milder constraint than those derived from the observation. 
With $\epsilon =-1$, however, (\ref{bd5-2c}) implies
\beq
\xi^{-1}< 6,\quad{\rm or}\quad \omega < \frac{3}{2},
\label{bd5-13b}
\eeq
which is ruled out immediately by the solar-system experiments, giving $\omega\gsim 10^3$ [\cite{will}]. We nevertheless continue our analysis as long as theoretical consistency is maintained.

From (\ref{bd5-11}) also follows
\beq
\phi(t) =\phi_{0}t^{1/2},
\label{bd5-15}
\eeq 
where $\phi_{0}$ is determined by identifying (\ref{bd5-6a}) with
(\ref{bd5-12a}); $\phi_{0}^4=16 C_{1}r^2 m_{0}^4 \beta^2 \xi^{-2}$,
which, if used in (\ref{bd5-3}),  yields
\beq
m(t)= \half \left( C_{1}r^2 \beta^2 \right)^{-1/4}t^{-1/2}.
\label{bd5-16}
\eeq
It is interesting to notice that the behavior $m(t)\sim t^{-1/2}$ follows simply because the potential should be proportional to $m^{4}$, as shown in (\ref{bd5-6}), and it must decay like $t^{-2}$ because it is part of the energy density appearing on the right-hand side of the 00-component of the Einstein equation.

Obviously the assumption $r=\:$const is crucial in the above argument.  We could obtain $m=\:$const if $r(t)\sim t^{-1}$, but with a highly unreasonable consequence that $r$ should be as large as $\sim 10^{63}$ at the Planck time.  Also the dependence $V\sim m^4$ is common to any diagrams of many $\phi^{-1}$'s, as in (b) and (c) in Fig. 1.  Including them results simply in affecting the overall size of the potential.

We arrived at (\ref{bd5-16}) in E frame in which the standard technique of quantum field theory can be applied.  We should also notice, however, that we use some type of microscopic 
clocks in most of the measurements. The time unit of atomic clocks, for example, is provided typically by the frequency $\sim m\alpha^4.$  
It is also important to recognize that the cosmic time is usually assumed to be measured in the same time unit.
If the time unit $\tau(t)$ itself changes with time, the new time $\tilde{t}$ measured in this unit would be defined by
\beq
d\tilde{t}=\frac{dt}{\tau(t)}.
\label{bd5-20}
\eeq
In conformity with special relativity, the scale factor $a(t)$ in Robertson-Walker cosmology is transformed in the same manner:
\beq
\tilde{a}=\frac{a(t)}{\tau(t)}.
\label{bd4-3}
\eeq
These two relations can be combined to a conformal transformation
\beq
d\tilde{s}^2 =\tau^{-2}ds^2, \quad{\rm or}\quad
g_{\mu\nu}=\tau^2 \tilde{g}_{\mu\nu}.
\label{bd4-4}
\eeq

In the prototype BD model, there is no mechanism to make $\alpha$
time-dependent, hence $\tau\sim m^{-1}$.  Combining this with $m\sim
\phi^{-1}$ as derived from (\ref{bd5-3}), and also comparing
(\ref{bd4-4}) 
with (\ref{bd55_5}), we find that (\ref{bd4-4}) implies going back
to the original J frame, as it should because it is the frame in which mass $m_{0}$ is taken to be constant.

We now try to solve the cosmological equations in J frame, in which we also suppress tildes to simplify the notation. It is also interesting to find that the behavior $V_{1}\sim \phi^{-4}$ as indicated in (\ref{bd5-6}) shows that this potential in E frame can be derived from a cosmological constant in J frame as given by  
\beq
\Lambda =C_{1}r^2 m_{0}^4.
\label{bd8_4}
\eeq
{\em The quantum effect computed in E frame amounts to introducing $\Lambda$ back in the original conformal frame.}  The field equations in J frame are given by 
\beqa
2\varphi G_{\mu\nu}&=& T_{\mu\nu}+T_{\mu\nu}^{\phi}-g_{\mu\nu}\Lambda
-2\left( g_{\mu\nu}\BBox -\nabla_{\mu}\nabla_{\nu}  \right)\varphi,
\label{bd8_6}\\
\BBox\varphi &=& \beta^2 (T -4\Lambda),\label{bd8_6a}\\
\nabla_{\mu}T^{\mu\nu}&=& 0.\label{bd8_7}
\eeqa
Notice also that $T_{\mu\nu}$ is the matter energy-momentum tensor while
\beq
T_{\mu\nu}^{\phi}=\epsilon\left( \partial_{\mu}\phi\partial_{\nu}\phi
-\half g_{\mu\nu}\left( \partial \phi  \right)^2\right) .
\label{bd8_8}
\eeq
Assuming spatially uniform $\phi$, we derive the cosmological equations:
\beqa
6\varphi H^2&=& \epsilon\half \dot{\phi}^2 +\Lambda +\rho -6 H\dot{\varphi},\label{bd8_9z}\\
\ddot{\varphi}+3H\dot{\varphi}&=&4\beta^2  \Lambda,\label{bd8_9}\\
\dot{\rho} +4 H{\rho}&=&0,\label{bd8_10}
\eeqa
where we have chosen $T=0$ confining ourselves to the radiation-dominated era even if $\Lambda$ comes from nonzero $m_{0}$.

As a heuristic approach, let us choose
\beq
H=0.
\label{bd8_11}
\eeq
Then (\ref{bd8_10}) leads to 
\beq
\rho = {\rm const}.
\label{bd8_12}
\eeq
Using (\ref{bd8_11}) in (\ref{bd8_9}) we obtain $\ddot{\varphi}= 4\beta^2 \Lambda,$ which allows a solution
\beq
\varphi(t)=2 \beta^2 \Lambda t^2 +\varphi_{1}t +\varphi_{0}.
\label{bd8_14}
\eeq
Notice that we have chosen $\Lambda >0$ hence $C_{1}>0$ so that
$\sigma$ falls off the potential slope toward infinity in E frame. This implies that
$\varphi$ increases also in J frame as indicated in (\ref{bd5-2}) if
$\beta^2 >0$.  This is the very condition, however, which is in contradiction with the observation, as already discussed in connection with (\ref{bd5-13b}).  Taking aside this drawback for the moment again, we expect
\beq
\phi\approx \sqrt{\frac{4\Lambda}{6\xi -1}}\,t,
\label{bd8_15}
\eeq
at sufficiently late times.  Using this together with (\ref{bd8_11}) and (\ref{bd8_12}) in (\ref{bd8_9z}) gives
\beq
\rho = 3\Lambda \frac{1-2\xi}{6\xi -1},
\label{bd8_15a}
\eeq
which is positive if (\ref{bd5-13a}) and (\ref{bd5-13b}) are obeyed.

An example is shown in Fig. 2, in which we see how $H$ approaches zero, much faster than $t^{-1}$; the Universe quickly becomes stationary after alternate occurrences of expansion ($H>0$) and contraction ($H<0$). This together with other similar examples indicate strongly that there is an ``attractor" to
which solutions of different initial conditions would approach
asymptotically.  Fig. 3, which is a 2-dimensional cross section of the 3-dimensional phase space of $\varphi,\;\dot{\varphi}$ and $\rho$, illustrates how different solutions are attracted to a common destination given by \reflef{bd8_15}) and \reflef{bd8_15a}), which represent in fact a curve in the whole phase space as one finds because of the relation $\dot{\phi}^2=\dot{\varphi}^2/(2\xi\varphi)$.  A trajectory for a set of initial values proceeds along, spiraling around and coming ever closer to this curve.

One might be tempted to compare our solutions with those in Einstein's model with a negative cosmological constant $\Lambda <0$, but of course without $\phi$.  This model allows a static solution $\rho = -\Lambda$ and $H=0$, but the Universe would never become stationary in contrast to our solutions; $\rho$ with a sufficiently large initial value decreases toward the minimum $-\Lambda$ but bounces back to increase in a touch-and-go fashion.

In this way we come to conclude that the BD model corrected for an important quantum effect should result in a steady state Universe, which is totally unacceptable in view of the success of the standard model in understanding primordial nucleosynthesis.

The constant scale factor in J frame may be interpreted also from the analysis in E frame, in which the length unit is provided by $\tau\sim\phi\sim
t^{1/2}$ which increases {\em in the same rate} as $a\sim t^{1/2}$ shown in
(\ref{bd5-10}) [\cite{nth}]. We also find that $\phi\sim t$ as given by
(\ref{bd8_15}) in J frame and $\phi\sim t_{*}^{1/2}$ in
(\ref{bd5-15}), in which the symbol $*$ is restored in E frame, are
consistent with each other, since $t\sim t_{*}^{1/2}$ is a consequence of the relation (\ref{bd5-20}).\footnote{Apply the replacement,
$\tilde{t}\rightarrow t, t\rightarrow t_{*}.$    For the dust-dominated Universe with $a_{*}\sim t_{*}^{2/3}$, we find $a\sim t^{1/6}.$}  These observations seem
to support (\ref{bd5-9}) which is only approximate unlike its counter
part (\ref{bd8_10}) in J frame.

On the other hand, one may ask if there is any sensible solution with
 $a\sim t^{1/2}$ in J frame.  In (\ref{bd8_10}) we substitute 
\beq
H=(1/2)t^{-1},
\label{bd10_5}
\eeq
thus obtaining 
\beq
\rho =\rho_{0}t^{-2}.
\label{bd10_6}
\eeq
Then (\ref{bd8_9}) becomes
\beq
\ddot{\varphi}+\frac{3}{2}t^{-1}\dot{\varphi} =4\beta^2 \Lambda,
\label{bd10_7}
\eeq
which is solved asymptotically:
\beq
\varphi\approx \frac{4}{5}\beta^2 \Lambda t^2, \quad
{\rm or}\quad \phi\approx \sqrt{\epsilon\frac{8}{5}\left( 1
-6\beta^2  \right) \Lambda }\,t.
\label{bd10_8}
\eeq
We then find that the right-hand side of (\ref{bd8_9z}) is given by
\beq
\rho +\frac{3}{5}\left( 3-16\beta^2 \right)\Lambda.
\label{bd10_9}
\eeq

Now from (\ref{bd10_5}) and (\ref{bd10_8}), the left-hand side of
(\ref{bd8_9z}) should be time-independent.  This can be matched with the situation in which $\rho$ given by (\ref{bd10_6}) decreases rapidly to be negligible compared with the second term of (\ref{bd10_9}); implying that the Universe becomes asymptotically ``vacuum dominated,'' again an unrealistic conclusion.  Even
worse, ignoring $\rho$ in the right-hand side of (\ref{bd10_9}) and
using (\ref{bd10_8}) on the left-hand side of (\ref{bd8_9z}) yields
\beq
\beta^2 =\frac{1}{6},
\label{bd10_10}
\eeq
which on substituting into (\ref{bd5-2a}) gives
\beq
\xi^{-1}=0,
\label{bd10_11}
\eeq
hardly a realistic result.\footnote{The same analysis applied to the dust matter results in $\epsilon =-1$ and $\xi =-2/3$, being inconsistent with (\ref{bd55_5a}).}

\section{Discussions}
We add that our argument of choosing J frame is independent of whether we literally use atomic clocks to measure something during 
the epoch in question.  It is simply in accordance with realistic situations that analyses are based on quantum mechanics in which mass of every particle is taken to be truly constant.

We admit that there should be some other quantum effects to be included.  The result obtained here is, however, so remote from what would be expected from the standard scenario that it is highly unlikely that those ``other" effects conspire miraculously to restore the success in the nucleosynthesis, among other things.  It seems that we need some more fundamental modification of the model.

A possible way out is to abandon the assumption (iv) about the absence of the {\em direct} $\phi$ coupling to the matter in J frame. As an extreme counter example, we may replace the matter Lagrangian (\ref{bd55_4})  in J frame by
\beq
L_{\psi}=-\psibar\left( \Dslash +f\phi \right)\psi,
\label{bd5-33}
\eeq
where $f$ is a dimensionless Yukawa coupling constant.  $\psi$ is massless in J frame, while in E frame we obtain the mass $m= f\xi^{-1/2}$ (in units of $(8\pi G)^{-1/2}$) which is {\em independent of} $\sigma$.  The scalar field is {\em decoupled} from $\psi$ {\em in E frame}, hence is left {\em invisible} through the matter coupling; it plays a role {\em only in cosmology}, most likely as a form of dark matter. With {\em constant} mass in E frame, which is now physical, we may reasonably adjust parameters such that the standard scenario is reproduced to a good approximation.

There might be intermediate choices between the prototype model and this extreme model.  Then we may expect the matter coupling generically {\em much weaker} than
\beq
H'_{\sigma_{1}}=-\beta m(t)\psibar\psi\sigma,
\label{bd5-30}
\eeq
in the prototype model, hence evading immediate conflicts with the
test of WEP and the constraint from the solar-system
experiments.\footnote{See Ref. [\cite{fn}] for a model of this type.}  Needless to say, $G$ is predicted to be constant, by construction.

With modifications of this nature in mind, we add a comment on the mass term of the scalar field, which is ought to arise from $V_{1}$ as given by (\ref{bd5-6}). We should be interested here in a {\em fluctuating component} $\sigma_{1}(x)$ which is responsible for the force between local mass distributions, to be separated from the spatially uniform component $\sigma_{0}(t)$ evolving as the cosmic time $t$;
\beq
\sigma(x)=\sigma_{0}(t)+\sigma_{1}(x),
\label{bd5-24}
\eeq
which satisfies
\beq
\BBox\sigma -V'(\sigma)=0,
\label{bd5-25}
\eeq
from which (\ref{bd5-8}) derives.

The cosmological component $\sigma_{0}$ is a solution of (\ref{bd5-25}) with $\sigma_{1}$ dropped;
\beq
\BBox\sigma_{0}+4\beta V_{1}(0)e^{-4\beta\sigma_{0}}=0.
\label{bd5-26}
\eeq
If we use this in (\ref{bd5-25}) for the entire field (\ref{bd5-24}), we obtain
\beq
\BBox\sigma_{1}-4\beta V_{1}(0)e^{-4\beta\sigma_{0}}
\left( 1-e^{-4\beta\sigma_{1}} \right)=0.
\label{bd5-27}
\eeq
Expanding the terms in the last parenthesis, we find
\beq
\BBox\sigma_{1}=\mu^2\sigma_{1} +\cdots,
\label{bd5-28}
\eeq
where, using (\ref{bd5-11}) for $\sigma_{0}(t)$,\footnote{For dust-dominated Universe, the right-hand side is doubled.}
\beq
\mu^2 =16\beta^2 V_{1}(0)t^{-2},
\label{bd5-29}
\eeq
which is $V''(\sigma)$ at $\sigma = \sigma_{0}$.

This shows that the scalar field does acquire a ``mass" even though the potential has {\em no stationary point}, but the range of the force mediated by $\sigma_{1}$ is basically given by $t$, which is the size of the visible part of the Universe at each epoch.  The force-range at the present epoch can be as ``short" as $10^5$ ly if $16\beta^2 V_{1}(0)\sim 10^{10}$, in contrast to (\ref{bd5-12a}).  It is rather likely that the force can be considered to be infinite-range in any practical use.

We may relax the assumption $F\sim \phi^2$ in the prototype model.  We recognize, however, that the relation of the type $V(\sigma)\sim m^4(\sigma)$ as in (\ref{bd5-6}) is quite generic and so is $m\sim t^{-1/2}$ according to the argument following (\ref{bd5-16}).  This makes the conclusion (\ref{bd8_11}) almost inevitable, as long as the assumption (iv) is maintained.

As another aspect of more general $F(\phi)$, we point out that the factor $\xi^{-1/2}\phi^{-1}$ in (\ref{bd5-1}) is in fact $F^{-1/2}$.  It then follows that the potential as given by (\ref{bd5-6}) generalizes to
\beq
V_{1}(\sigma)\sim m^{4}\sim F^{-2}.
\label{bd5-34}
\eeq
The relation (\ref{bd5-2}) is also traced back to
\beq
\frac{d\sigma}{d\phi}= F^{-1}\sqrt{\epsilon F +\frac{3}{2}F'\:^2}.
\label{bd5-35}
\eeq
We may then expect that the potential as a function of $\sigma$ would have a minimum if $F(\phi)$ has a maximum, the same conclusion as in Refs. [\cite{dm}].

The potential minimum should act, however, as an effective cosmological constant, which might present another serious conflict with observations unless it remains below the level of $\sim t^{-2}_{0}\sim 10^{-120}$ at the present epoch.  In this respect we have an advantage in the potential having no stationary point.

As the last comment we point out that the occurrence of a ghost which
was required to give positive matter density is not entirely unnatural from the point of view of unified theories.  Consider, for example, Kaluza-Klein approach to $4+n$ dimensional spacetime.  The size of compactified {$n$}-dimensional space behaves as a 4-dimensional scalar field, which is shown to have {\em wrong sign} in the kinetic term.  This model provides also one of the natural origins of the nonminimal coupling.

\begin{center}
{\Large\bf Acknowledgments}
\end{center}
I thank Akira Tomimatsu and Kei-ichi Maeda for valuable discussions.

\begin{center}
{\Large\bf References}
\end{center}
\begin{enumerate}
\item\label{bd}C. Brans and R.H. Dicke, Phys. Rev. {\bf 124}(1961)925.\item\label{yf1}Y. Fujii, Mod. Phys. Lett. {\bf A9}(1994)3685.
\item\label{dicke}R.H. Dicke, Phys. Rev. {\bf 125}(1962)2163.
\item\label{fn}Y. Fujii and T. Nishioka, Phys. Rev. {\bf D42}(1990)361.
\item\label{cho}Y.M. Cho, Phys. Rev. Lett. {\bf 68}(1992)3133.
\item\label{jd}P. Jordan, {\sl Schwerkraft und Weltalle}, (Friedrich
Vieweg und Sohn, Braunschweig, 1955).\item\label{will}See, for example, C. Will, {\sl Theory and Experiment in Gravitational Physics,} rev. ed., Cambridge University Press, Cambridge,1993.
\item\label{nth}T. Nishioka, Thesis, University of Tokyo, 1991.
\item\label{dm}T. Damour and K. Nordtvedt, Phys. Rev. Lett. {\bf 70}(1993)2217; T. Damour and A.M. Polyakov, Nucl. Phys. {\bf B423}(1994)532.
\end{enumerate}

\newpage
%%%%%%%%%%%%%%%%%%%%%%%%%%%%%%%%%%%%%%%%%%%%%%

\epsfverbosetrue

\begin{figure}[h]
\hspace*{1cm}
\epsfxsize=12cm
\epsfbox{abc4.eps}
\caption{Some of the one-$\psi$-loop diagrams, shown only up to the third order in $\phi^{-1}$.  Each blob represents $\phi^{-1}$
appearing in (12), a collection of many vertices of $\sigma$'s, while a solid line is for  the $\psi$ line.} 
\end{figure}

\begin{figure}[h]
\centering
\hspace*{1cm}
\epsfxsize=12cm
\epsfbox{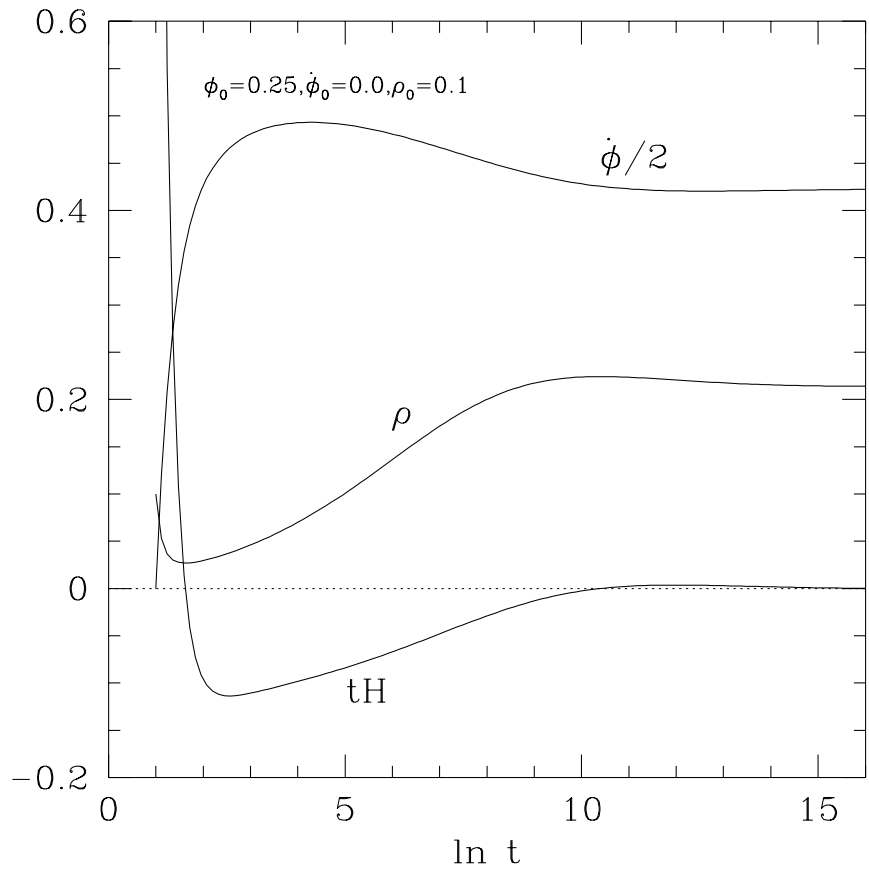}
\caption{A solution of (38)-(40) with the initial values $\varphi_{0}=0.25.\; \dot{\varphi}_{0}=0.0,\; \rho_{0}=0.1$ at $\ln t=1$.  We chose $\Lambda = 0.5$ and $\xi =0.4$.}
\end{figure}

\begin{figure}[h]
\hspace{2cm}
\epsfxsize=10cm
\epsfbox{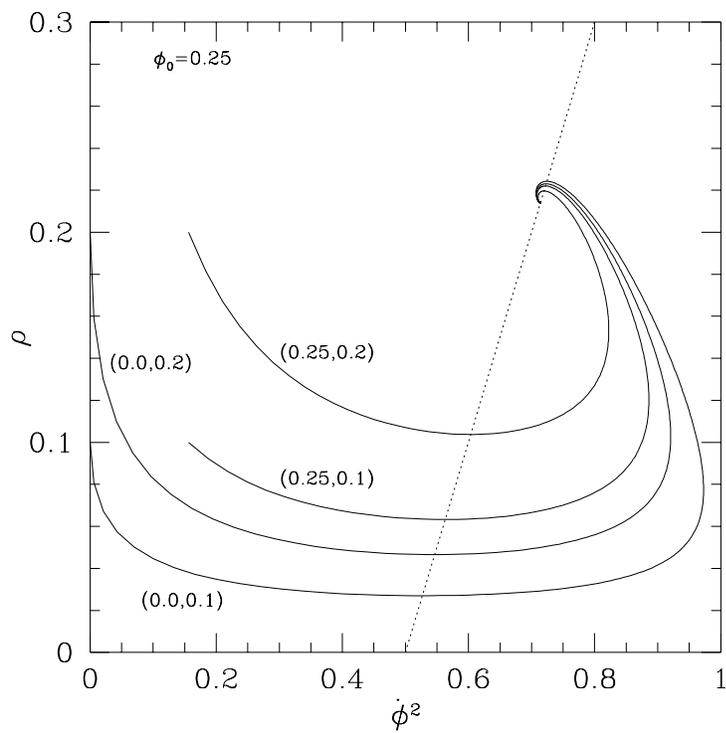}
\caption{Trajectories ($\ln t=1-30$) of the solutions of different initial values $\dot{\varphi}_{0}$ and $\rho_{0}$, as shown in the parentheses, with other parameters the same as in Fig. 2. The point of convergence ($\dot{\phi}^2=1.42857, \rho =0.21429$) is given by (44) and (45).  The dotted line is for $H=0$; its left-side for $H>0$.} 
\end{figure}

\end{document}